\documentclass[aoas,MSNbibl,nameyear,seceqn,dvips]{arximspdf}
\usepackage{algorithm,algorithmic}
\usepackage{dcolumn,multirow}
\usepackage{graphicx}

\doi{10.1214/14-AOAS771} 
\volume{8}
\issue{4}
\pubyear{2014}
\firstpage{2292}
\lastpage{2318}
\docsubty{FLA}

\makeatletter
\newcolumntype{d}[1]{D{.}{.}{#1}}
\newcommand{\rrvert}{\vert}

\newcommand{\llvert}{\vert}

\newtheorem{teo}{Theorem}
\newproclaim{rem}{Remark}
\newcommand{\bgamma}{\bolds{\gamma}}
\makeatother

\begin{document}
\begin{frontmatter}

\title{A fast algorithm for detecting gene--gene interactions in
genome-wide association studies}
\runtitle{Detect gene--gene interactions in GWAS}

\begin{aug}
\author[A]{\fnms{Jiahan} \snm{Li}\ead[label=e1]{jli7@nd.edu}\thanksref{m1}},
\author[B]{\fnms{Wei} \snm{Zhong}\corref{}\thanksref{m2,T1}\ead[label=e2]{wzhong@xmu.edu.cn}},
\author[C]{\fnms{Runze} \snm{Li}\thanksref{m3,T2}\ead[label=e3]{rli@stat.psu.edu}}
\and
\author[D]{\fnms{Rongling} \snm{Wu}\ead[label=e4]{rwu@hes.hmc.psu.edu}\thanksref{m3}}
\runauthor{Li, Zhong, Li and Wu}
\affiliation{University of Notre Dame\thanksmark{m1},
Xiamen University\thanksmark{m2}\\ and
Pennsylvania State University\thanksmark{m3}}
\address[A]{J. Li\\
Department of Applied and Computational\\
\quad Mathematics and Statistics\\
University of Notre Dame\\
Notre Dame, Indiana 46556\\
USA\\
\printead{e1}}
\address[B]{W. Zhong\\
Wang Yanan Institute\\
\quad for Studies in Economics\hspace*{20pt}\\
Department of Statistics\\
School of Economics\\
Fujian Key Laboratory\\
\quad of Statistical Science\\
Xiamen University\\
Xiamen, Fujian 361005\\
China\\
\printead{e2}}
\address[C]{R. Li\\
The Methodology Center\\
Department of Statistics\\
Pennsylvania State University\\
University Park, Pennsylvania 16802\\
USA\\
\printead{e3}}
\address[D]{R. Wu\\
Center for Statistical Genetics \\
Pennsylvania State University \\
Hershey, Pennsylvania 17033\\
USA\\
\printead{e4}}
\end{aug}
%
\thankstext{T1}{Supported by NNSFC Grants 11301435, 71131008 and the Fundamental Research Funds for the Central Universities 20720140034.
Wei Zhong is the corresponding author.}
\thankstext{T2}{Supported by NIDA Grant P50-DA10075 and NNSFC Grant 11028103.}

\received{\smonth{7} \syear{2012}}
\revised{\smonth{5} \syear{2014}}

%
\begin{abstract}
With the recent advent of high-throughput genotyping techniques,
genetic data for genome-wide association studies (GWAS) have become
increasingly available, which entails the development of efficient and
effective statistical approaches. Although many such approaches have
been developed and used to identify single-nucleotide polymorphisms
(SNPs) that are associated with complex traits or diseases, few are
able to detect gene--gene interactions among different SNPs. Genetic
interactions, {also known as epistasis}, have been recognized to play a
pivotal role in contributing to the genetic variation of phenotypic
traits. However, because of an extremely large number of SNP--SNP
combinations in GWAS, the model dimensionality can quickly become so
overwhelming that no prevailing variable selection methods are capable
of handling this problem. In this paper, we present a statistical
framework for characterizing main genetic effects and epistatic
interactions in a GWAS study. Specifically, we first propose a
two-stage sure independence screening (TS-SIS) procedure and generate a
pool of candidate SNPs and interactions, which serve as predictors to
explain and predict the phenotypes of a complex trait. {We also propose
a rates adjusted thresholding estimation (RATE) approach to determine
the size of the reduced model selected by an independence screening.}
Regularization regression methods, such as LASSO or SCAD, are then
applied to further identify important genetic effects. Simulation
studies show that the TS-SIS procedure is computationally efficient and
has an outstanding finite sample performance in selecting potential
SNPs as well as gene--gene interactions. We apply the proposed framework
to analyze an ultrahigh-dimensional GWAS data set from the Framingham
Heart Study, and select 23 active SNPs and 24 active epistatic
interactions for the body mass index variation. It shows the capability
of our procedure to resolve the complexity of genetic control.
\end{abstract}

%
\begin{keyword}
\kwd{Gene--gene interaction}
\kwd{GWAS}
\kwd{high-dimensional data}
\kwd{sure independence screening}
\kwd{variable selection}
\end{keyword}
\end{frontmatter}

\setcounter{footnote}{2}

\section{Introduction}\label{sec1}
Genome-wide association studies (GWAS) have been a powerful tool for
genetic and biomedical research. The past decade has witnessed the
rapid development of GWAS and the substantial contributions it has made
[\citet{AltDalLan08}; \citet{Psy09}; \citet{Hir09};
\citet{Dasetal11}]. With advances in high-throughput genotyping
techniques and modern statistics, GWAS have been helping investigators
understand the genetic basis of many complex traits or diseases,
providing valuable clues to the genetic predisposition of common
diseases and drug responses {[\citet{Con07}; \citet{Dal10}]},
among others.


In a typical GWAS, hundreds of thousands of single-nucleotide
polymorphisms (SNPs) are usually genotyped on a cohort being studied to
identify important genetic variants that are associated with the trait
of interest. Although fast and inexpensive, the collection of genetic
information is normally limited to a sample involving {hundreds of
subjects}, which brings statistical challenges for estimating and
identifying relevant genetic risk factors. With SNPs being predictors
and phenotypes being the response, single-SNP analysis is mostly
performed. However, such a single-SNP approach is neither efficient nor
precise, since it fails to consider all SNPs and their possible
interactions simultaneously, and to adjust the estimated effects
accordingly. Therefore, many statistical procedures that consider all
SNPs jointly have been proposed for analyzing the high-dimensional data
sets generated by genome-wide association studies.

This is a feature selection problem for high-dimensional data, where
the number of SNPs ($p$) is much larger than the number of observations
($n$). Penalized regressions, which are developed to overcome severe
drawbacks of traditional variable selection techniques, are widely used
to select a subset of important predictors from a large number of
potential predictors. In the GWAS analysis of case-control studies, \citet{Wuetal09} and \citet{Choetal09} applied LASSO penalized regression
[\citet{Tib96}] and elastic-net penalized regression [\citet{ZouHas05}], respectively. \citet{AyeCor10} further
conducted a comprehensive study to examine the performance of a variety
of penalized regressions in case-control studies. They concluded that
variable selection techniques based on penalized regressions outperform
single-SNP analysis and stepwise selection. {To further explore the
potential of high-dimensional statistical models for identifying
disease susceptibility genes}, several two-stage approaches have been
proposed for selecting significant main effects. \citet{Lietal11}
employed preconditioning and Bayesian LASSO on population cohorts to
estimate genetic effects of SNPs on continuous traits. \citet{HeLin11} developed a GWASelect procedure for case-control cohorts, where
several steps of iterative sure independence screening (ISIS) and LASSO
regression are involved.

These methods based on penalized regressions have demonstrated their
statistical power and computational feasibility over the single-SNP
analysis. However, since statistical methodologies and computations
have already been challenged by the overwhelming number of whole-genome
SNPs, these methods either do not consider gene--gene interactions or
estimate interactions among only a small number of selected SNPs with
significant main effects. However, without considering\vadjust{\goodbreak} the full picture
of epistatic interactions in GWAS analysis, only a limited portion of
phenotypic variation can be explained such that potential
disease-associated pathways and risk factors can hardly be identified
[\citet{Manetal09}; \citet{Cor09}].

In light of recent developments in machine learning, many sophisticated
approaches have been proposed to search for whole-genome interactions
in genome-wide association studies, most of which are designed for
case-control cohorts. These machine learning approaches include a
Bayesian partitioning model [\citet{ZhaLiu07}], a SNPRuler based on
an association rule [\citet{Wanetal10N2}] and random forest approaches
[{\citet{Bre01};} \citet{Kimetal09}]. However, these methods are
computationally intensive and do not perform well in practice {when
the} genome-wide SNP data {are} considered [\citet{Wanetal10N2}; \citet{Wanetal11}; \citet{Szyetal09}]. More recently, adaptive LASSO
[\citet{Yanetal10}] and Bayesian generalized linear models [\citet{YiKakPas11}] are applied to detect epistatic
interactions in case-control cohorts where all SNP pairs are
exhaustively searched. \citet{Wanetal11} present a comprehensive
comparison of the prevailing epistatic interaction detection methods,
including SNPRuler [\citet{Wanetal10N2}], SNPHarvester [\citet{Yanetal09}], Screen and Clean [\citet{Wuetal10}], BOOST [\citet{Wanetal10N1}] and TEAM [\citet{Zhaetal10}]. They concluded that these
methods perform differently in terms of statistical power, false
positive rate and computational cost. However, methods other than
Screen and Clean are specially designed for case-control studies where
phenotypic values are binary, and cannot be applied to {quantitative
traits} unless the phenotypical values are properly discretized.

In this paper, we propose a statistical framework for detecting
whole-genome epistatic interactions in a population cohort where
phenotype is continuous. The framework incorporates the well-developed
penalized regressions, which have proved successful in detecting SNPs
with significant main effects. Therefore, existing findings regarding
penalized regression theories and their empirical performance in GWAS
analysis can provide direct and valuable insights into our framework.
{Moreover, the proposed algorithm is suitable for parallel computing
and does not involve computationally demanding techniques on the
whole-genome SNP data that prevailing interaction models may involve,
such as resampling strategies and Bayesian analysis. As a result, it is
computationally efficient.}

Specifically, we develop a two-stage sure independence screening
(TS-SIS) procedure before variable selection. The screening step forms
a pool of important SNPs, which may either have significant main
effects or demonstrate no marginal effects but strong epistatic
interactions. Since the two-stage screening is based on sure
independence screening [\citet{FanLv08}], the computational burden of
selecting important interactions is greatly reduced. More importantly,
this procedure guarantees the performance of the following variable
selection procedure, in the sense that once important SNPs and
interactions enter the pool, the probability of identifying the correct
ones is very high. We also propose a rates adjusted thresholding
estimation (RATE) approach to determine the number of predictors
retained by a variable screening procedure. This approach is based on
soft-thresholding and bootstrapping, and relates the reduced model size
to a false positive rate. {\citet{UekTam12} proposed
hard-thresholding-based sure independence screening (SIS) to select
promising main genetic effects and interactions for penalized
regressions. Motivated by the multifactor dimensionality reduction
[MDR; \citet{Ritetal01}] approach, they proposed dummy coding
methods to effectively capture various patterns of interactions in
case-control studies. Our approach, however, is more general and
suitable for population-based GWAS and other variable screening problems.}

We applied the newly developed statistical framework to analyze {a GWAS
data set} from the Framingham Heart Study, aimed to identify genetic
variants that are associated with obesity, blood pressure and heart
disease. We find that, out of 349,985 SNPs, 23 SNPs and 24 epistatic
interactions have notable effects on the body mass index (BMI). {By
applying gene-set enrichment analysis tools [\citet{WanLiBuc07};
\citet{Holetal08}] in future studies, biological knowledge can be
integrated to discover and prioritize signaling pathways implied by
detected SNPs. Morever, SNP--SNP interactions will provide insight into
functional related genes and the structure of genetic pathways,
allowing better understanding of complex genetic architecture and
cellular processes in a system level. }


In Section~\ref{sec2} we introduce the TS-SIS procedure that reduces the model
dimensionality and identifies potential gene--gene interactions.
Section~\ref{sec3} proposes a rates adjusted thresholding estimation (RATE)
approach to determine the number of predictors retained by a general
variable screening procedure. Section~\ref{sec4} shows how penalized regression
can fit in this framework and gives the estimation procedure for SCAD
penalized regression [\citet{FanLi01}]. In Section~\ref{sec5} the statistical
properties of this framework are investigated through simulation
studies. Section~\ref{sec6} applies this framework to the Framingham Heart Study.
Concluding remarks are given in Section~\ref{sec7}.

\section{Two-stage sure independence screening}\label{sec2}
In genome-wide association studies phenotypical measurements are
explained by a handful of covariates and a great number of genetic
factors represented by SNP genotypes. To select important SNPs and
estimate their genetic effects precisely by adjusting for observed
covariates, we employ a GWAS model that takes into account the effects
of both genetic effects and covariate effects. Moreover, as we
discussed, epistatic interactions play a central role in understanding
metabolic pathways of complex diseases and traits. Therefore, a
comprehensive GWAS model incorporating both main genetic effects and
gene--gene interactions is more appropriate.

For subject $i$ in a population cohort consisting of a total of $n$
subjects, we describe the observed phenotypic value $y_i$ as
%
\begin{eqnarray}\label{mod1}\label{e2.1}
y_i &=& \mu+ \sum
_{k=1}^{q} x_{k,i} \alpha_k +
\sum_{j=1}^{p} \xi_{j,i}
a_j + \sum_{j=1}^{p}
\zeta_{j,i} \,d_j + \sum_{j=1}^{p}
\sum_{j' < j} \xi_{j,i} \xi_{j',i}
\mathcal{I}_{j
j'}^{aa}\nonumber
\\
&&{} + \sum_{j=1}^{p} \sum
_{j'=1}^{p} \xi_{j,i}
\zeta_{j',i} \mathcal {I}_{j j'}^{ad} + \sum
_{j=1}^{p} \sum_{j'=1}^{p}
\zeta_{j,i} \xi_{j',i} \mathcal {I}_{j j'}^{da}
\\
&&{}
+ \sum_{j=1}^{p} \sum
_{j' < j} \zeta_{j,i} \zeta_{j',i} \mathcal
{I}_{j j'}^{dd} + \varepsilon_i,\nonumber
\end{eqnarray}
where $\mu$ is the overall mean, $q$ is the number of nongenetic
covariates, $p$ is the number of SNPs, $x_{k,i}$ is the $k$th covariate
for subject $i, k = 1, \ldots,q, i = 1,\ldots,n$, which could be
either discrete or continuous, $\alpha_k$ is the effect of the $k$th
covariate, $a_j$ and $d_j$ are the additive effect and dominant effect
of the $j$th SNP, respectively, for $j = 1, \ldots, p$, $\mathcal
{I}_{j j'}^{aa}$ is the additive${}\times{}$additive epistatic effect
between the $j$th SNP and the $j'$th SNP, $\mathcal{I}_{j j'}^{ad}$,
$\mathcal{I}_{j j}^{da}$ and $\mathcal{I}_{j j'}^{dd}$ are additive${}\times{}$dominant epistatic\vspace*{1pt} effect, dominant${}\times{}$additive
epistatic effect and dominant${}\times{}$dominant epistatic effect, and
$\varepsilon_i$ is the residual error assumed to follow a $N(0,\sigma
^2)$ distribution.
{If an effect is nonzero in the regression model (\ref{mod1}), we say
that the corresponding covariate or interaction is active.}
For subject $i$, $\xi_{j,i}$ and $\zeta_{j,i}$ are the indicators of
the additive and dominant effects of the $j$th SNP, respectively, which
are defined as
\begin{eqnarray*}
\xi_{j,i} &=& \cases{ 1, &\quad if the genotype of SNP $j$ is $AA$,
\vspace*{2pt}\cr
0, &\quad if the genotype of SNP $j$ is $Aa$,
\vspace*{2pt}\cr
-1,&\quad if the genotype of SNP
$j$ is $aa$,}
\\
\zeta_{j,i} &=& \cases{ 1, &\quad if the genotype of SNP $j$ is $Aa$,
\vspace*{2pt}\cr
0, &\quad if the genotype of SNP $j$ is $AA$ or $aa$.}
\end{eqnarray*}
{Therefore, the additive effect $a_j$ in model (\ref{e2.1}) measures the
change of the average phenotypic value by substituting allele $A$ with
allele $a$ in a population. Dominant effect $d_j$, on the other hand,
represents how the effect of allele $A$ is modified by the presence of
allele $a$, allowing a more general nonadditive genetic model.}

Given observed phenotypic traits, genetic information and covariates
such as gender or age, our goal is to characterize the genetic control
of the phenotype, by selecting active SNPs and gene--gene interactions
and estimating their genetic effects. However, since in GWAS data sets,
the number of SNPs usually far exceeds the number of subjects, it is
almost impossible to directly estimate all genetic effects, as even
epistatic interactions are not considered in the regression model.
Recently, penalized regressions that regularize the size of regression
coefficients are applied to GWAS models without interactions, and
appropriate algorithms are designed for high-dimensional inference,
such as cyclical coordinate descent methods. But in many clinical
trials where the number of SNPs is extremely large compared with the
sample size, the empirical performance of penalized regression is not
guaranteed. Moreover, if four interaction terms for each SNP pair are
considered in the GWAS analysis, the estimation of all genetic effects
in the ultrahigh-dimensional setting is infeasible from the perspective
of both statistical theories and computational cost.

To identify this ultrahigh-dimensional model in practice, and to make
the best use of GWAS data for better explanation and predictions, we
need to put assumptions on the heredity structures of epistatic
effects, although we want to make the restrictions as weak as possible.
Two versions of the effect heredity principle are the following: strong
heredity and weak heredity [\citet{Chi96}]. Under strong heredity, if
the interaction between two predictors is significant, both predictors
should be marginally significant. Under weak heredity, only one needs
to be marginally significant.

Obviously, in prevailing penalized regression models for GWAS, where
interaction effects are tested after a subset of significant SNPs are
selected, strong heredity assumption is implicitly imposed. However,
throughout this paper we will {assume only} weak heredity, since, in
practice, many important SNPs are marginally uncorrelated with the
response, {but interact with other SNPs} in an epistasis network. With
this biologically meaningful assumption in the epistatic GWAS model as
well as large data sets collected in genome-wide studies, the potential
of GWAS could be fully explored, and a detailed picture of genetic
control and regulation could be unveiled.

Two SNPs involved in a two-way interaction will be denoted as ``two roots.''
We will employ a two-stage sure independence screening (TS-SIS)
procedure to identify SNPs which may have active main effects or may
act as roots. Sure independence screening is a statistical learning
technique for ultrahigh-dimensional data proposed by \citet{FanLv08}.
In the context of GWAS analysis, it ranks the importance of SNPs
according to their marginal correlations with the response and retains
those SNPs whose marginal correlations are strong enough. It can be
shown that under some technical conditions, sure independence screening
enjoys the sure screening property. That is, the reduced model is
capable of retaining all the active SNPs with asymptotic probability one.

Let $\mathcal{D}_a$ and $\mathcal{D}_d$ be two sets of indices of
truly important additive effects and truly important dominant effects,
respectively. {The first SIS round will be performed between each SNP
and the response to select active main effects. Since it is common
practice to include covariates as linear predictors of the response in
GWAS analysis, covariates are not subject to SIS and will later be
added to the reduced model after TS-SIS.} After the first stage of SIS,
two subsets of SNPs with potential nonzero additive effects
$\widehat{\mathcal{D}}_a$ and potential\vspace*{1pt} nonzero dominant effects
$\widehat{\mathcal{D}}_d$ are selected. Sure\vspace*{1pt} screening property [\citet{FanLv08}] implies that truly important main effects are retained in
$\widehat{\mathcal{D}}_a$ and $\widehat{\mathcal{D}}_d$ with high
probabilities.

Next, we formulate pairwise epistatic interactions between all SNPs in $\widehat{\mathcal{D}}_a$ or
$\widehat{\mathcal{D}}_d$ and all genome-wide SNPs.
In particular, an additive$\times$additive interaction term is formulated by taking one
SNP from $\widehat{\mathcal{D}}_a$ and taking any additive effect from all SNPs. The set of
additive${}\times{}$additive interactions are denoted by $\mathcal{D}_{aa}^{(0)}=\{(j,j')\dvtx
\xi_{j} \xi_{j'}, j\in \widehat{\mathcal{D}}_a, j'=1,2,\ldots,p\}$. Similarly,
additive${}\times{}$dominant interactions $\mathcal{D}_{ad}^{(0)}$, dominant${}\times{}$additive
interactions $\mathcal{D}_{da}^{(0)}$, and dominant${}\times{}$dominant interactions
$\mathcal{D}_{dd}^{(0)}$ are formulated, and the GWAS model becomes
\begin{eqnarray}\label{mod2}
\quad y_i &=&
\mu + \sum_{k=1}^{q} x_{k,i} \alpha_k
+ \sum_{j \in \widehat{\mathcal{D}}_a} \xi_{j,i} a_j
+ \sum_{j \in \widehat{\mathcal{D}}_d} \zeta_{j,i} d_j
+ \sum_{(j,j') \in \mathcal{D}_{aa}^{(0)}}  \xi_{j,i} \xi_{j',i} \mathcal{I}_{j j'}^{aa}\nonumber
\\
&&{}
+ \sum_{(j,j') \in \mathcal{D}_{ad}^{(0)}} \xi_{j,i} \zeta_{j',i} \mathcal{I}_{j j'}^{ad}
+ \sum_{(j,j') \in \mathcal{D}_{da}^{(0)}} \zeta_{j,i} \xi_{j',i} \mathcal{I}_{j j'}^{da}
\\
&&{}
+ \sum_{(j,j') \in \mathcal{D}_{dd}^{(0)}} \zeta_{j,i} \zeta_{j',i} \mathcal{I}_{j j'}^{dd}
+ \varepsilon_i.\nonumber
\end{eqnarray}

After adding interaction terms in the model (\ref{mod2}), the model dimensionality becomes
extremely high compared with GWAS model without epistatic interactions.
To test whether these interactions contribute to the observed variation in
phenotypes, we again apply SIS to all interaction terms and select epistatic
effects that are highly correlated with the response. Let $\widehat{\mathcal{D}}_{aa}$ be the index
set for the selected additive${}\times{}$additive interactions between a
SNP in $\widehat{\mathcal{D}}_a$ and another genome-wide SNP.
Similarly, we define three other sets, $\widehat{\mathcal{D}}_{ad}$, $\widehat{\mathcal{D}}_{da}$
and $\widehat{\mathcal{D}}_{dd}$, which contain selected  additive${}\times{}$dominant,
dominant${}\times{}$additive and dominant$\times$dominant interactions, respectively. Then the GWAS model after TS-SIS becomes
\begin{eqnarray}\label{e2.3}
y_i &=&
\mu + \sum_{k=1}^{q} x_{k,i} \alpha_k
+ \sum_{j \in \widehat{\mathcal{D}}_a} \xi_{j,i} a_j
+ \sum_{j \in \widehat{\mathcal{D}}_d} \zeta_{j,i} d_j
+ \sum_{(j,j') \in \widehat{\mathcal{D}}_{aa}} \xi_{j,i} \xi_{j',i} \mathcal{I}_{j j'}^{aa}\nonumber \\
&&{}
+ \sum_{(j,j') \in \widehat{\mathcal{D}}_{ad}} \xi_{j,i} \zeta_{j',i} \mathcal{I}_{j j'}^{ad}
+ \sum_{(j,j') \in \widehat{\mathcal{D}}_{da}} \zeta_{j,i} \xi_{j',i} \mathcal{I}_{j j'}^{da}
\\
&&{}+ \sum_{(j,j') \in \widehat{\mathcal{D}}_{dd}} \zeta_{j,i} \zeta_{j',i} \mathcal{I}_{j j'}^{dd}
+ \varepsilon_i.\nonumber
\end{eqnarray}

Algorithm~\ref{alg1} summarizes the TS-SIS procedure, where the sizes of the
reduced models in steps 1~and~3 will be determined by the RATE
approach proposed in Section~\ref{sec3}.

\begin{algorithm}
\caption{Two-stage sure independence screening}\label{alg1}
\begin{algorithmic}
\STATE{\textit{Step} 1.} Apply the SIS approach to all additive and
dominate main effects SNPs and estimate the reduced models $\widehat
{\mathcal{D}}_a$ and $\widehat{\mathcal{D}}_d$.
\STATE{\textit{Step} 2.}
Formulate pairwise epistatic interactions between all SNPs selected in $\widehat{\mathcal{D}}_a$ or
$\widehat{\mathcal{D}}_d$ and all genome-wide SNPs $\mathcal{D}=\{1,2,\ldots,p\}$. That is,\vspace*{1pt}
  $\mathcal{D}_{aa}^{(0)}=\{(j,j')\dvtx \xi_{j} \xi_{j'}, j\in \widehat{\mathcal{D}}_a, j'\in \mathcal{D}\}$,
  $\mathcal{D}_{ad}^{(0)}=\{(j,j')\dvtx \xi_{j} \zeta_{j'}, j\in \widehat{\mathcal{D}}_a, j'\in \mathcal{D}\}$,
  $\mathcal{D}_{da}^{(0)}=\{(j,j')\dvtx \zeta_{j} \xi_{j'}, j\in \widehat{\mathcal{D}}_d, j'\in \mathcal{D}\}$, and
  $\mathcal{D}_{dd}^{(0)}=\{(j,j')\dvtx \zeta_{j} \zeta_{j'}, j\in \widehat{\mathcal{D}}_d, j'\in \mathcal{D}\}$.
\STATE{\textit{Step} 3.} Apply\vspace*{1pt} the SIS approach again to all epistatic
interactions in step~2, that is, $\mathcal{D}_{aa}^{(0)}$, $\mathcal
{D}_{ad}^{(0)}$, $\mathcal{D}_{da}^{(0)}$
and $\mathcal{D}_{dd}^{(0)}$, and\vspace*{1pt} obtain the reduced models $\widehat
{\mathcal{D}}_{aa}$, $\widehat{\mathcal{D}}_{ad}$, $\widehat
{\mathcal{D}}_{da}$~and~$\widehat{\mathcal{D}}_{dd}$.
\STATE{\textit{Step} 4.} Combine all reduced models in steps~1~and~3 to
obtain the final selected model by the TS-SIS procedure: $\{ \widehat
{\mathcal{D}}_a, \widehat{\mathcal{D}}_d, \widehat{\mathcal
{D}}_{aa}, \widehat{\mathcal{D}}_{ad}, \widehat{\mathcal{D}}_{da},
\widehat{\mathcal{D}}_{dd}\}$.
\end{algorithmic}
\end{algorithm}\vspace*{-9pt}


\begin{rem*}
If some nongenetic covariates (such as age) are known as
truly significant predictors in model (\ref{e2.1}),
the following modified independence screening procedure can be
implemented to improve the performance in steps~1~and~3 of Algorithm~\ref{alg1}.
We run a linear regression of the response on each SNP and the
significant nongenetic covariates,
and utilize the magnitude of the SNP's estimated coefficient as a
marginal screening utility.\footnote{We thank the Associate Editor
for suggesting this modified independence screening.}
\end{rem*}

\section{Rates adjusted thresholding estimation}\label{sec3}
In this section we propose a general rule to determine the size of the
reduced model selected by an independence screening procedure. This
rule can be applied to other independence screening methods. In its
application to the\vspace*{1pt} proposed TS-SIS, it is equivalent to determining the
cardinalities of sets $\widehat{\mathcal{D}}_{a}$, $\widehat
{\mathcal{D}}_{d}$,
$\widehat{\mathcal{D}}_{aa}$, $\widehat{\mathcal{D}}_{ad}$,
$\widehat{\mathcal{D}}_{da}$ and $\widehat{\mathcal{D}}_{dd}$.

In general, the choice of the reduced model size is critical for any
independence screening approach. If the model size is too large, the
following penalized regression would be less efficient due to the
presence of too many noise variables. If the model size is too small,
on the other hand, it is likely to miss important predictors in the
screening stage. \citet{FanLv08} suggested the reduced model size
being proportional to $[n/\log n]$ for the SIS procedure, where $n$ is
the sample size and~$[\cdot]$ denotes the integer of a real number.
Although this hard thresholding is easy to implement in practice,
little theoretical evidence is provided to guarantee its performance in
different data sets. \citet{Zhuetal11} proposed a soft-thresholding
rule by adding auxiliary variables in their Sure Independent Ranking
and Screening (SIRS) procedure for multi-index models with
ultrahigh-dimensional covariates. In what follows we propose a general
data-driven procedure to determine the reduced model size that extends
the soft-thresholding procedure.

Denote the $p_n$-dimensional vector of predictors by $\mathbf
{x}=(X_1,\ldots,X_{p_n})$, and denote the vector of regression
coefficients by {$\bgamma=(\gamma_1,\ldots,\gamma_{p_n})$} in a
linear regression model. Let $\mathcal{M}$ be the set of {active
predictors} and $\mathcal{M}^c$ be its complement. That is, $\mathcal
{M}=\{1\leq j \leq p_n\dvtx  \gamma_j\neq0\}$ and $\mathcal{M}^c=\{1\leq
j \leq p_n\dvtx  \gamma_j= 0\}$. 
The idea of the soft-thresholding rule in \citet{Zhuetal11} is as
follows. First, $d$ auxiliary variables are generated independently and
randomly $\mathbf{z}=(Z_1,\ldots,Z_d)\sim N_d(\mathbf{0},\mathbf
{I}_d)$. Next, an\vspace*{1pt} independence screening procedure is applied to the
combined predictors set $(\mathbf{x}^{\mathrm{T}},\mathbf
{z}^{\mathrm{T}})^{\mathrm{T}}
$. Let $\rho_k$ be the marginal screening utility between each
predictor and the response, where $k=1,2,\ldots,(p_n+d)$. Because
$\mathbf{z}$ is known to be independent of the response, the marginal
utility $\rho_k$ between any $Z_k$ and the response is exactly zero
and the associated sample version $\hat{\rho}_k$ should be less than
any marginal utility between the {active predictors} and the response.
\citet{Zhuetal11} suggested the maximal sample marginal utility of all
auxiliary variables, $C_d = \max_{1 \leq m \leq d} \hat{\rho
}_{p_n+m}$, as a natural cutoff to separate two sets of active and
inactive predictors in $\mathbf{x}$. Thus, the selected model is
determined by $\widehat{\mathcal{M}}=\{1\leq j \leq p_n\dvtx  \hat{\rho
}_j > C_d\}$.

Although the soft-thresholding procedure may be useful, there are two
major concerns of practical interest. The first is the choice of the
number of auxiliary variables $d$. The larger the $d$ value, the
sparser the selected model, and thus the higher the probability of
missing some active predictors. Besides, a larger $d$ value implies
more computation cost. On the other hand, a smaller $d$ value gives a
smaller cutoff, thus the reduced model dimensionality could still be
very high. The second concern is how to generate independent auxiliary
variables $\mathbf{z}$. The performance of the soft-thresholding rule
depends on an exchangeability assumption between inactive predictors
and auxiliary variables assumed in Theorem~3 of \citet{Zhuetal11}. But
its validity is difficult to check in practice.\footnote{For instance,
both the Editor and Associate Editor mentioned that the distribution of
``noisy'' SNPs is quite different from a normal distribution, so the
exchangeability assumption may be violated. We thank the Editor and
Associate Editor for pointing this out.} To address these concerns, we
propose a rates adjusted thresholding estimation (RATE) approach to
determine the number of auxiliary variables $d$ by bootstrapping
auxiliary variables from the original data.

In particular, we propose to relate the number of auxiliary variables
$d$ to the false positive rate of an independence screening procedure
\[
\frac{\llvert \widehat{\mathcal{M}}\cap\mathcal{M}^c\rrvert }{\llvert \mathcal{M}^c\rrvert }, %
\]
which is the proportion of inactive predictors that are incorrectly
included in the selected model $\widehat{\mathcal{M}}$. {In data
mining and bioinformatics, statistical power is also known as
sensitivity, and false positive rate is one minus specificity. Both
sensitivity and specificity are performance measures of interest in
genetic association studies [see, e.g., \citet{Dugetal08};
\citet{Goretal08}; \citet{Haretal08}; \citet{Jacetal09}].}

The next theorem provides a lower bound on the probability that the
false positive rate is controlled under a pre-specified level $\alpha$.

\begin{teo}\label{teofpr}\label{teo1}
Suppose that the inactive variables $\{X_j\dvtx  j\in\mathcal{M}^c\}$ and
auxiliary variables $\{Z_k\dvtx  k=1,\ldots,d\}$ are exchangeable
in the sense that the\vadjust{\goodbreak} inactive and auxiliary variables are equally
likely to be selected by the independence screening procedure.
Under the sparsity condition that $s_n<n$, the probability that the
false positive rate can be controlled under a
pre-specified level $\alpha$ is bounded from below.
That is,
%
\begin{equation}
P \biggl(\frac{\llvert \widehat{\mathcal{M}}\cap\mathcal
{M}^c\rrvert }{\llvert \mathcal{M}^c\rrvert }< \alpha \biggr)\ge1- \biggl\{1-\frac{\alpha(p_n-n)}{p_n+d}
\biggr\}^{d}.
\end{equation}
\end{teo}

The theorem implies that the probability of the false positive rate
being controlled below a given level $\alpha$ is greater than $1-
 \{1-\frac{\alpha(p_n-n)}{p_n+d} \}^{d}$.
Given a fixed confidence level $1-\beta= 1- \{1-\frac{\alpha
(p_n-n)}{p_n+d} \}^{d}$, the number of auxiliary variables $d$
can be determined.
According to Theorem \ref{teofpr}, we propose the RATE procedure in
Algorithm~\ref{alg2} for a general independence screening method.

\begin{algorithm}[t]
\caption{Rates adjusted thresholding estimation}\label{algorate}\label{alg2}
\begin{algorithmic}
\STATE{\textit{Step} 1.} Solve the equation $ \{1-\frac{\alpha
(p_n-n)}{p_n+d} \}^{d}=\beta$ to obtain $d$, for given $p_n$,
$n$, $\alpha$ and $\beta$.
\STATE{\textit{Step} 2.} Bootstrap the original data $\mathbf
{x}=(X_1,\ldots,X_{p_n})$ to obtain $d$ independent auxiliary
variables $\mathbf{z}=(Z_1,\ldots,Z_{d})$ in the following way. For
each $i=1,\ldots,n$, randomly assign one value of $(X_{1k},\ldots,X_{(i-1)k},X_{(i+1)k},\ldots,X_{nk})$ to $Z_{ik}$, and then get
the vector $Z_k=(Z_{1k},\ldots,Z_{nk})$ for $k=1,\ldots,d$.
If $d>p_n$, one may iterate the above procedure until getting enough
auxiliary variables.
\STATE{\textit{Step} 3.} Compute the marginal screening utility $\hat{\rho
}^*_k$ between each auxiliary variable $Z_k$ and the response,
$k=1,\ldots,d$, and set the cutoff $C_d = \max_{1 \leq k \leq d} \hat
{\rho}^*_{k}$.
\STATE{\textit{Step} 4.} Compute the marginal screening utility $\hat{\rho
}_j$ between each predictor $X_j$ and the response, $j=1,\ldots,p_n$,
and select the reduced model as $\widehat{\mathcal{M}}=\{1\leq j \leq
p_n\dvtx  \hat{\rho}_j > C_d\}$.
\end{algorithmic}
\end{algorithm}

{We remark that the modified bootstrapping procedure in step~2 in
Algorithm~\ref{alg2} is to guarantee the independence between the response and
auxiliary variables $\mathbf{z}$.
We obtain independent auxiliary variables by bootstrapping the original
data instead of simulating them from a normal distribution.
Consequently, the bootstrapped auxiliary variables have the same data
structure as the original predictors, approximating the exchangeability
condition in the soft-thresholding rule.} Note that with given $p_n$
and $n$, two rates $\alpha$ and $\beta$ together determine the number
of auxiliary variables $d$. Therefore, we call this approach the rates
adjusted thresholding estimation (RATE). It will be shown later that
the RATE approach {has excellent performance} in the simulation studies
and the real data analysis.

\section{SCAD penalized regression}\label{sec4}
After two-stage sure independence screening, the dimensionality of the
GWAS model {is greatly reduced}. In order to precisely select important
SNPs and epistatic interactions from a pool of candidate \mbox{effects},
penalized regressions widely used in main-effect analysis could be\vadjust{\goodbreak}
incorporated here. Specifically, we put penalties on the sizes of
additive effects, dominant effects and all epistatic effects and
minimize the following penalized least squares:
%
\begin{eqnarray}\label{eqPSL}
\nonumber
\qquad&&\frac{1}{2n}\|\mathbf{y}-E{\bf y}\|^2
+\sum_{j \in {\widehat{\mathcal{D}}_a}} p_{\lambda}\bigl(|a_j|\bigr)
+\sum_{j \in {\widehat{\mathcal{D}}_d}} p_{\lambda}\bigl(|d_j|\bigr)
+\sum_{(j,j') \in \widehat{\mathcal{D}}_{aa}} p_{\lambda}\bigl(\bigl|\mathcal{I}_{j j'}^{aa}\bigr|\bigr)\nonumber
\nonumber\\[-8pt]\\[-8pt]\nonumber
&&\qquad{}+
\sum_{(j,j') \in {\widehat{\mathcal{D}}_{ad}}} p_{\lambda}\bigl(\bigl|\mathcal{I}_{j j'}^{ad}\bigr|\bigr)
+ \sum_{(j,j') \in {\widehat{\mathcal{D}}_{da}}} p_{\lambda}\bigl(\bigl|\mathcal{I}_{j j'}^{da}\bigr|\bigr)
+ \sum_{(j,j') \in {\widehat{\mathcal{D}}_{dd}}} p_{\lambda}\bigl(\bigl|\mathcal{I}_{j j'}^{dd}\bigr|\bigr),\nonumber
\end{eqnarray}
where the penalty function $p_{\lambda}(\cdot)$ is implemented to
shrink sufficiently small effects to zero and thus exclude
the inactive predictors.

We consider the smoothly clipped absolute deviation (SCAD) penalty
function due to its
unbiasedness, continuity and sparsity properties [\citet{FanLi01}].
The SCAD penalty is a nonconvex function and
defined as follows:
\begin{eqnarray*}
p_{\lambda}(b)&=&\lambda\llvert b\rrvert I\bigl(0\leq\llvert b\rrvert <
\lambda\bigr) + \frac{a\lambda\llvert  b\rrvert
-(b^2+\lambda^2)/2}{a-1} I\bigl(\lambda\leq\llvert b\rrvert \leq a\lambda
\bigr)
\\[-3pt]
&&{} +\frac
{(a+1)\lambda^2}{2}I\bigl(\llvert b\rrvert >a\lambda\bigr),
\end{eqnarray*}
where $I(\cdot)$ is an indicator function and $a=3.7$ as suggested in
\citet{FanLi01}. $\lambda$ is the tuning parameter which balances
the model complexity and forecasting performance. We follow the idea of
\citet{WanLiTsa07} and choose $\lambda$ by a BIC tuning
parameter selector.


Commonly-used algorithms for the SCAD penalized least squares include
the local quadratic approximation (LQA) algorithm [\citet{FanLi01}],
the perturbed LQA [\citet{HunLi05}] and the local linear
approximation (LLA) [\citet{ZouLi08}] algorithm. With the aid of LLA,
one may employ the LARS algorithm to obtain the SCAD estimate. Thus, we
will use the LLA algorithm in this paper. Specifically, for a given
initial value $\beta^{(0)}$, the penalty function $p_{\lambda}(\cdot
)$ can be locally approximated by a linear function as
%
\begin{equation}
\qquad p_{\lambda}\bigl(\llvert \beta\rrvert \bigr) \approx p_{\lambda
}\bigl(
\bigl\llvert \beta^{(0)}\bigr\rrvert \bigr)+p_{\lambda}^{\prime}
\bigl(\bigl\llvert \beta^{(0)}\bigr\rrvert \bigr) \bigl(\llvert \beta
\rrvert -\bigl\llvert \beta^{(0)}\bigr\rrvert \bigr) \qquad\mbox{for }
\llvert \beta\rrvert \approx\bigl\llvert \beta^{(0)}\bigr\rrvert.
\end{equation}
With the aid of LLA, the estimates of regression coefficients in SCAD
penalized least squares (\ref{eqPSL}) can be obtained by minimizing
%
\begin{eqnarray}\label{eqPSLLLA}
&&\frac{1}{2n}\|\mathbf{y}-E\mathbf{y}\|^2
+\sum_{j \in {\widehat{\mathcal{D}}_a}} p^{\prime}_{\lambda}\bigl(\bigl|a_j^{(0)}\bigr|\bigr)|a_j|
+\sum_{j \in {\widehat{\mathcal{D}}_d}} p^{\prime}_{\lambda}\bigl(\bigl|d_j^{(0)}\bigr|\bigr)|d_j|\nonumber
\\
&&\qquad{}+\sum_{(j,j') \in \widehat{\mathcal{D}}_{aa}} p^{\prime}_{\lambda}\bigl(\bigl|\mathcal{I}_{j j'}^{aa(0)}\bigr|\bigr)
\bigl(\bigl|\mathcal{I}_{j j'}^{aa}\bigr|\bigr)
+\sum_{(j,j') \in {\widehat{\mathcal{D}}_{ad}}} p^{\prime}_{\lambda}\bigl(\bigl|\mathcal{I}_{j j'}^{ad(0)}\bigr|\bigr)
\bigl(\bigl|\mathcal{I}_{j j'}^{ad}\bigr|\bigr)
\\
&&\qquad{}+\sum_{(j,j') \in {\widehat{\mathcal{D}}_{da}}} p^{\prime}_{\lambda}\bigl(\bigl|\mathcal{I}_{j j'}^{da(0)}\bigr|\bigr)\bigl(\bigl|\mathcal{I}_{j j'}^{da}\bigr|\bigr)
+\sum_{(j,j') \in {\widehat{\mathcal{D}}_{dd}}} p^{\prime}_{\lambda}\bigl(\bigl|\mathcal{I}_{j j'}^{dd(0)}\bigr|\bigr)
\bigl(\bigl|\mathcal{I}_{j j'}^{dd}\bigr|\bigr),\nonumber
\end{eqnarray}
after constants are discarded. Note that this penalized least squares
can be easily minimized based on $L_1$ penalized regression.

\section{Simulated studies}\label{sec5}
In this section we investigate the GWAS analysis framework consisting
of TS-SIS and variable selection through simulation studies. We
simulate large data sets where SNPs may {have either (a) main effects
or (b) interaction effects}. Our goal is to identify these active SNPs
with high accuracy and low computational cost.

%
\begin{table}[b]
\tabcolsep=0pt
\tablewidth=270pt
\caption{Information of 6 assumed genetic effects for data simulation}\label{tab1}
\begin{tabular*}{\tablewidth}{@{\extracolsep{\fill}}@{}lcccc@{}}
\hline
\textbf{Chr.} & \textbf{Position} & \textbf{Additive/dominant} & \textbf{Interact with} & \textbf{Effect size}\\
\hline
\multicolumn{5}{@{}c@{}}{\textit{Main effects}}\\
\phantom{0}1 & 1 & Additive & -- & 1 \\
\phantom{0}2 & 1 & Dominant & -- & 1 \\
\phantom{0}3 & 1 & Additive & -- & 1
\\[3pt]
\multicolumn{5}{@{}c@{}}{\textit{Epistatic interactions}}\\
11 & 1 & Additive & 1 & 1 \\
\phantom{0}2 & 2 & Dominant & 2 & 1 \\
12 & 1 & Dominant & 2 & 1 \\
\hline
\end{tabular*}
\end{table}

Specifically, genotypes of $p$ SNPs across 23 chromosomes are generated
for $n = 500$ subjects. For SNP $j$ of subject $i, j = 1, \ldots, p, i
= 1, \ldots, n$, its genotype $\xi_{j,i}$ is derived from $u_{j,i}$,
{where the vector $(u_{1,i},\ldots,u_{p,i})$ is generated from
multivariate normal distribution with zero mean and covariance matrix
$\Sigma=(\sigma_{j,k})_{p\times p}$, $\sigma_{j,k} = \rho^{\llvert  j-k\rrvert }$ for $\rho= 0.2$, $0.5$ or $0.8$.} Then, we set
\[
\xi_{j,i} = \cases{ 1, &\quad$u_{ij} > c_{1j}$,
\vspace*{2pt}\cr
0, &\quad$c_{2j} \le u_{ij} \le c_{1j}$,
\vspace*{2pt}\cr
-1, &
\quad$u_{ij} < c_{2j}$,}
\]
{where $c_{1j}$ and $c_{2j}$ determine the minor allele frequency
(MAFs). We consider two cases: homogeneous case, MAF${}=0.5$ for each $j$,
and heterogeneous case, in which
the MAF of each SNP is randomly set to 0.5, 0.35 or 0.2 with equal
likelihood.} Finally, the dominant effect indicator $\zeta_{j,i}$ is
derived from $\xi_{j,i}$ by setting $\zeta_{j,i} = 1-\llvert \xi
_{j,i}\rrvert $. In total, there are $p = 3948$ SNPs across 23
chromosomes, with the number of SNPs in each chromosome being one
percent of that in a real data set we are going to work on.

We put 3 active main effects and 3 active epistatic interactions across
the whole genome, whose positions and effect sizes are given in
Table~\ref{tab1}. Column ``Interact with'' in Table~\ref{tab1} indicates, out of 3 active
main effects, which one the SNP interacts with. When simulating the
response variable, we standardize the design matrix columnwisely, such
that all columns of the design matrix have the same variance. This step
makes the comparison of detecting active main effects and active
interactions fair. From Table~\ref{tab1}, it can be seen that one SNP could
interact with two other SNPs without marginal effects (three SNPs on
chromosomes 2 and 12), and two SNPs involved in a two-way interaction
may also be correlated (two SNPs on chromosomes 2). These interaction
patterns add further complexity in the simulation studies.

For each simulated data set, we first implement SIS with the RATE
procedure to select $s_1$ SNPs, which may exhibit notable main effects
or epistatic effects. We determine $s_1$ in each simulation according
to Theorem~\ref{teo1} with $\alpha=0.01$ and $\beta=0.0001$ and, on average,
there are 11 SNPs selected in the first stage of TS-SIS.
According to the sure screening property, this subset of $s_1$ SNPs
should include the first SNPs on the first 3 chromosomes with high
probability, which demonstrate active main effects and may serve as
roots in two-way interactions.

To select those SNPs that have no marginal effects, but modify the
genetic effects of other SNPs, two-way interactions are formed between
each selected SNP in the first stage and any SNP across the genome
according to model (2). SIS is carried out again, and, in total, $s_2$
pairs of SNPs are selected. We set $\alpha=0.005$ and $\beta=0.0001$.
These SNP pairs should contain all epistatic interactions, although
they may rank low in terms of the absolute value of marginal
correlations. Finally, $s_1$ SNPs with potential main effects and $s_2$
SNP pairs enter model (\ref{e2.3}), and variable selections are implemented to
select important SNPs and estimate their main effects and epistatic
effects. We consider both LASSO regression and SCAD regression
following TS-SIS.

Table~\ref{tab2} reports the statistical power, false positive rates and
computational time using R code. The result is the average over {100
simulations} with standard error in parenthesis. In columns labeled
``TS-SIS'' under ``Power~(\%),'' we present the statistical power of
TS-SIS, or the proportion of active SNPs and interactions that are
successfully included in the candidate pool of $s_1$ SNPs and $s_2$
interactions. In adjacent columns ``TS-SIS-SCAD'' and ``TS-SIS-LASSO,''
we report statistical powers of two-stage SIS paired with SCAD
regression or LASSO regression, or the percent of 6 active SNPs that
are correctly identified by the whole procedure. Note that the
statistical power under ``TS-SIS-SCAD'' or ``TS-SIS-LASSO'' cannot be
greater than that of TS-SIS, since a SNP or an epistatic interaction is
considered by the variable selection procedure only if it is correctly
identified by TS-SIS. In each column under ``False Positive Rate
($\times$10$^{-4}$),'' we report the false positive rate defined as the
proportion of unimportant SNPs that are incorrectly identified. We also
report the median computing time for TS-SIS with penalized regression
over all replications. The simulation is conducted on a 32-bit windows
7 system, with an Intel (R) i5-2400 processor, 3.10 GHz, 4G memory.

%
\begin{table}
\tabcolsep=0pt
\caption{Statistical power, false positive rate and running time of the proposed TS-SIS approach}\label{powerFPR}\label{tab2}
\begin{tabular*}{\tablewidth}{@{\extracolsep{\fill}}@{}ld{2.3}d{3.3}d{3.3}d{2.3}d{2.3}d{2.3}d{2.2}@{}}
\hline
& \multicolumn{3}{c}{\textbf{Power (\%)}} & \multicolumn{3}{c}{\textbf{False positive rate $\bolds{(\times10^{-4})}$}} &\\[-6pt]
& \multicolumn{3}{c}{\hrulefill} & \multicolumn{3}{c}{\hrulefill}\\
&                                                               & \multicolumn{1}{c}{\textbf{TS-SIS-}} & \multicolumn{1}{c}{\textbf{TS-SIS-}}&                                    &  \multicolumn{1}{c}{\textbf{TS-SIS-}} &\multicolumn{1}{c}{\textbf{TS-SIS-}}& \multicolumn{1}{c@{}}{\textbf{Time}}\\
$\bolds{(\rho,\sigma^2)}$ & \multicolumn{1}{c}{\textbf{TS-SIS}} & \multicolumn{1}{c}{\textbf{SCAD}}    & \multicolumn{1}{c}{\textbf{LASSO}}  &\multicolumn{1}{c}{\textbf{TS-SIS}} &  \multicolumn{1}{c}{\textbf{SCAD}}    &\multicolumn{1}{c}{\textbf{LASSO}}& \multicolumn{1}{c@{}}{\textbf{(seconds)}}\\
\hline
\multicolumn{8}{@{}c@{}}{\textit{Homogeneous case} (\textit{MAF}${} = 0.50$)}\\
(0.8,~6) & 97.0 & 92.7 & 86.5 & 4.22 & 2.18 & 2.92 & 12.29 \\
& (8.3) & (11.7) & (13.3) & (2.35) & (1.09) & (1.47) & \\
(0.8,~8) & 95.5 & 86.2 & 82.3 & 4.52 & 2.42 & 3.19 & 13.64 \\
& (9.1) & (13.4) & (15.9) & (2.51) & (1.23) & (1.68) &
\\[3pt]
(0.5,~6) & 97.8 & 94.5 & 88.2 & 2.93 & 1.74 & 2.29 & 10.05 \\
& (6.6) & (9.5) & (13.2) & (1.58) & (0.88) & (1.16) & \\
(0.5,~8) & 95.8 & 90.7 & 88.5 & 2.76 & 1.72 & 2.21 & 11.13 \\
& (8.3) & (13.0) & (11.6) & (1.69) & (1.01) & (1.33) &
\\[3pt]
(0.2,~6) & 95.9 & 91.8 & 87.7 & 2.89 & 1.75 & 2.27 & 7.88 \\
& (9.0) & (11.7) & (13.4) & (1.88) & (1.05) & (1.43) & \\
(0.2,~8) & 95.5 & 87.8 & 87.1 & 2.70 & 1.81 & 2.28 & 9.12 \\
& (9.7) & (13.6) & (13.1) & (1.71) & (1.08) & (1.42) &
\\[6pt]
\multicolumn{8}{@{}c@{}}{\textit{Heterogeneous case} (\textit{mixed MAFs})} \\
(0.8,~6) & 98.17 & 89.83 & 89.50 & 4.71 & 2.30 & 3.18 & 12.41 \\
& (5.24) & (11.58) & (11.28) & (3.51) & (1.30) & (1.99) & \\
(0.8,~8) & 96.50 & 85.50 & 88.33 & 5.02 & 2.46 & 3.42 & 13.15 \\
& (7.22) & (12.46) & (11.73) & (4.17) & (1.49) & (2.11) &
\\[3pt]
(0.5,~6) & 98.17 & 91.50 & 90.67 & 3.47 & 1.87 & 2.45 & 11.17 \\
& (5.24) & (11.48) & (11.68) & (2.93) & (1.26) & (1.78) & \\
(0.5,~8) & 95.00 & 87.67 & 88.33 & 3.08 & 1.82 & 2.42 & 10.38 \\
& (9.91) & (13.94) & (14.11) & (2.36) & (1.12) & (1.75) &
\\[3pt]
(0.2,~6) & 97.67 & 90.17 & 90.83 & 3.08 & 1.80 & 2.37 & 10.59 \\
& (6.28) & (12.10) & (11.93) & (2.15) & (1.10) & (1.54) & \\
(0.2,~8) & 94.33 & 87.50 & 89.33 & 2.58 & 1.53 & 2.05 & 9.25 \\
& (9.54) & (13.06) & (11.24) & (2.33) & (0.99) & (1.61) & \\
\hline
\end{tabular*}
\end{table}

%
\begin{table}
\tabcolsep=0pt
\caption{Statistical power of detecting interactions of the proposed TS-SIS approach}\label{powerinteraction}\label{tab3}
\begin{tabular*}{\tablewidth}{@{\extracolsep{\fill}}@{}ld{3.2}d{3.2}d{3.2}d{3.2}d{3.2}d{3.2}@{}}
\hline
& \multicolumn{3}{c}{\textbf{Homogeneous case}} & \multicolumn{3}{c@{}}{\textbf{Heterogeneous case}}\\[-6pt]
& \multicolumn{3}{c}{\hrulefill} & \multicolumn{3}{c@{}}{\hrulefill}
\\
&                                                               & \multicolumn{1}{c}{\textbf{TS-SIS-}} & \multicolumn{1}{c}{\textbf{TS-SIS-}}&                                    &  \multicolumn{1}{c}{\textbf{TS-SIS-}} &\multicolumn{1}{c@{}}{\textbf{TS-SIS-}}\\
$\bolds{(\rho,\sigma^2)}$ & \multicolumn{1}{c}{\textbf{TS-SIS}} & \multicolumn{1}{c}{\textbf{SCAD}}    & \multicolumn{1}{c}{\textbf{LASSO}}  &\multicolumn{1}{c}{\textbf{TS-SIS}} &  \multicolumn{1}{c}{\textbf{SCAD}}    &\multicolumn{1}{c@{}}{\textbf{LASSO}}\\
\hline
(0.8,~6) & 93.3 & 93.3 & 92.7 & 97.0 & 97.0 & 97.0 \\
& (13.4) & (13.4) & (14.7) & (9.6) & (9.6) & (9.6) \\
(0.8,~8) & 96.7 & 96.3 & 96.3 & 94.7 & 94.3 & 94.7 \\
& (10.1) & (10.5) & (10.5) & (13.2) & (14.3) & (13.2)
\\[3pt]
(0.5,~6) & 96.7 & 96.7 & 96.7 & 96.7 & 96.7 & 96.7 \\
& (10.1) & (10.1) & (10.1) & (10.1) & (10.1) & (10.1) \\
(0.5,~8) & 92.6 & 92.6 & 92.6 & 91.0 & 91.0 & 91.0 \\
& (15.1) & (15.1) & (15.1) & (18.3) & (18.3) & (18.3)
\\[3pt]
(0.2,~6) & 93.5 & 93.5 & 93.5 & 96.0 & 96.0 & 96.0 \\
& (13.3) & (13.3) & (13.3) & (10.9) & (10.9) & (10.9) \\
(0.2,~8) & 92.7 & 92.7 & 92.7 & 91.0 & 91.0 & 91.0 \\
& (16.1) & (16.1) & (16.1) & (17.0) & (17.0) & (17.0) \\
\hline
\end{tabular*}
\end{table}

According to Table~\ref{tab2}, the TS-SIS captures most of the SNPs with active
main effects, as well as SNPs without main effects but demonstrating
active interactions. As a result, important SNPs are selected in the
reduced model, and the majority of irrelevant SNPs are eliminated
before variable selection. This critical step greatly improves the
probability of effectively identifying important SNPs and interactions
in GWAS analysis. After TS-SIS, SNPs and interactions in the reduced
model are selected by either SCAD or LASSO. As expected, variable
selection further reduces the false positive rate and increases the
interpretability of the final model.
{Compared with LASSO, SCAD can identify truly important SNPs with
higher probability for the homogeneous case.
As more SNPs have lower MAFs in the heterogeneous case, two penalized
regressions have comparable statistical powers.
In addition, SCAD delivers smaller false positive rates consistently in
all simulation scenarios.}
Table~\ref{tab2} further suggests that, as $\sigma^2$ decreases, the
statistical powers increase, but the linkage disequilibrium of two SNPs
measured by $\rho$ plays a limited role in this setting. Besides, this
variable screening procedure is very fast even though millions of
potential pairwise interactions are present in each simulation.

Table~\ref{tab3} gives the statistical power of detecting interactions, or the
average proportion of interactions that are selected over 100
simulations. By comparing Table~\ref{tab3} with Table~\ref{tab2}, it can be seen that
interactions are relatively more difficult to capture by variable
screenings than main effects. This is understandable since the number
of interaction terms is huge compared with the number of main effect
terms. Once important interactions are identified by TS-SIS, however,
they are unlikely to be missed by the following penalized regression.
As a result, the statistical power of the entire procedure is very
close to that of TS-SIS. In Table~\ref{tab4} we report the results when the
number of SNPs is doubled ($p=6996$) for MAF${} = 0.50$, with all other
specifications unchanged. Interestingly, although the statistical power
of TS-SIS increases, the power of SCAD and LASSO regressions slightly
decreases, because the same $\alpha$ and $\beta$ in Theorem~\ref{teo1} imply a
larger reduced model from \mbox{TS-SIS}.\footnote{On average, the total
number of main effects and interactions selected by TS-SIS increases
from 46.9 to 66.2.} However, given that the number of interactions
increases from about 24.5 million to about 98 million, the performance
of TS-SIS is excellent, as can be seen from the increased statistical
power and decreased false positive rates. In Table~\ref{tab4} we do not change
$\alpha$ and $\beta$ for comparison purposes; we
consider in future research the effects of user-specified rates.

%
\begin{table}
\tabcolsep=0pt
\caption{Statistical power, false positive rate and running time of
the proposed TS-SIS approach when the number of SNPs is doubled
($p=6996$) for MAF${}=0.5$}\label{tab4}
\begin{tabular*}{\tablewidth}{@{\extracolsep{\fill}}@{}ld{2.2}d{3.2}d{3.2}d{2.3}d{2.3}d{2.3}d{2.2}@{}}
\hline
& \multicolumn{3}{c}{\textbf{Power (\%)}} & \multicolumn{3}{c}{\textbf{False positive rate $\bolds{(\times10^{-4})}$}} &\\[-6pt]
& \multicolumn{3}{c}{\hrulefill} & \multicolumn{3}{c}{\hrulefill}\\
&                                                               & \multicolumn{1}{c}{\textbf{TS-SIS-}} & \multicolumn{1}{c}{\textbf{TS-SIS-}}&                                    &  \multicolumn{1}{c}{\textbf{TS-SIS-}} &\multicolumn{1}{c}{\textbf{TS-SIS-}}& \multicolumn{1}{c@{}}{\textbf{Time}}\\
$\bolds{(\rho,\sigma^2)}$ & \multicolumn{1}{c}{\textbf{TS-SIS}} & \multicolumn{1}{c}{\textbf{SCAD}}    & \multicolumn{1}{c}{\textbf{LASSO}}  &\multicolumn{1}{c}{\textbf{TS-SIS}} &  \multicolumn{1}{c}{\textbf{SCAD}}    &\multicolumn{1}{c}{\textbf{LASSO}}& \multicolumn{1}{c@{}}{\textbf{(seconds)}}\\
\hline
(0.8,~6) & 99.5 & 90.0 & 83.7 & 1.39 & 0.74 & 0.97 & 39.37 \\
& (3.7) & (12.9) & (14.2) & (0.64) & (0.29) & (0.40) & \\
(0.8,~8) & 97.3 & 87.8 & 84.1 & 1.32 & 0.78 & 1.01 & 44.68 \\
& (7.8) & (14.3) & (12.7) & (0.60) & (0.29) & (0.39) &
\\[3pt]
(0.5,~6) & 97.3 & 89.8 & 85.0 & 1.15 & 0.68 & 0.89 & 29.90 \\
& (7.0) & (11.8) & (14.3) & (0.58) & (0.32) & (0.41) & \\
(0.5,~8) & 97.1 & 86.3 & 83.7 & 1.17 & 0.74 & 0.95 & 36.56 \\
& (8.2) & (14.8) & (14.9) & (0.53) & (0.33) & (0.42) &
\\[3pt]
(0.2,~6) & 98.8 & 93.8 & 87.2 & 1.06 & 0.67 & 0.86 & 25.97 \\
& (4.3) & (10.2) & (12.5) & (0.47) & (0.28) & (0.36) & \\
(0.2,~8) & 94.7 & 80.7 & 84.0 & 1.18 & 0.77 & 0.99 & 33.52 \\
& (9.7) & (15.1) & (13.8) & (0.52) & (0.29) & (0.38) & \\
\hline
\end{tabular*}
\end{table}

We also compare this framework with other methods for detecting
SNP--SNP interactions in simulation studies with MAF${}={}$0.5. Although most
of the available interaction detection methods are designed for binary
phenotypes, the Mendel software program [\citeauthor{Lanetal01} (\citeyear{Lanetal01,Lanetal13})] and
the Screen and Clean (SC) method [\citet{Wuetal10}] can identify
important SNPs as well as interactions in GWAS analysis for the
quantitative phenotype. Moreover, they are scalable and computationally
efficient. Specifically, Analysis Option 24 in Mendel software is very
convenient to test for main genetic effects and interaction effects
based on marginal p-values or LASSO type analysis [\citet{WuLan08};
\citet{Wuetal09}; \citet{Zhoetal10}]. Table~\ref{tab5} reports the results from
four major analysis options of Mendel: (1) marginal analysis for main
effects followed by testing important marginal effects against all SNPs
for interactions (Mendel~1), (2)~marginal analysis for main effects
followed by testing all pairwise interactions among top SNPs (Mendel~2), (3) LASSO analysis for main effects followed by testing important
marginal effects against all SNPs for interactions (Mendel~3), and (4)
LASSO analysis for main effects followed by testing all pairwise
interactions among top SNPs (Mendel~4). Since these four analysis
options generate final models with pre-determined sizes, we use the
default model size of 10 for main effects and then determine the number
of selected interactions in a way that the final model size is the same
as our method (TS-SIS-SCAD). Table~\ref{tab5} also reports the performance of
the Screen and Clean (SC) method (column ``SC'') and
hard-thresholding-based TS-SIS (column ``Hard-SCAD'' and column
``Hard-LASSO''), where the first $[n/\log n]$ SNPs are selected in
TS-SIS. Since the final model size of Mendel is user specified, the
false positive rate is not reported.

%
\begin{table}
\tabcolsep=0pt
\caption{Statistical power and false positive rate of alternative methods}\label{tab5}
\begin{tabular*}{\tablewidth}{@{\extracolsep{\fill}}@{}ld{3.2}d{3.2}d{3.2}d{3.2}d{2.2}d{3.2}d{2.2}d{2.2}d{2.2}d{2.2}@{}}
\hline
& \multicolumn{7}{c}{\textbf{Power (\%)}} & \multicolumn{3}{c@{}}{\textbf{FPR $\bolds{(\times10^{-4})}$}}\\[-6pt]
& \multicolumn{7}{c}{\hrulefill} & \multicolumn{3}{c@{}}{\hrulefill}\\
& \multicolumn{1}{c}{\textbf{Hard-}} & \multicolumn{1}{c}{\textbf{Hard-}}
    && \multicolumn{1}{c}{\textbf{Mendel}}& \multicolumn{1}{c}{\textbf{Mendel}} & \multicolumn{1}{c}{\textbf{Mendel}}& \multicolumn{1}{c}{\textbf{Mendel}}
    & \multicolumn{1}{c}{\textbf{Hard-}} & \multicolumn{1}{c}{\textbf{Hard-}}
\\
$\bolds{(\rho,\sigma^2)}$ & \multicolumn{1}{c}{\textbf{SCAD}} &\multicolumn{1}{c}{\textbf{LASSO}}
    &\multicolumn{1}{c}{\textbf{SC}} &\multicolumn{1}{c}{\textbf{1}}&\multicolumn{1}{c}{\textbf{2}}& \multicolumn{1}{c}{\textbf{3}} & \multicolumn{1}{c}{\textbf{4}}
    & \multicolumn{1}{c}{\textbf{SCAD}} & \multicolumn{1}{c}{\textbf{LASSO}} & \multicolumn{1}{c@{}}{\textbf{SC}} \\
\hline
(0.8,~6) & 89.3 & 81.0 & 69.9 & 85.3 & 49.0 & 89.3 & 49.3 & 4.0 & 4.4 & 4.8 \\
& (9.3) & (8.5) & (10.5) & (8.7) & (4.0) & (10) & (3.3) & (0.4) & (0.4) & (2.7) \\
(0.8,~8) & 81.5 & 80.5 & 61.9 & 85.7 & 49.7 & 88.0 & 49.0 & 4.2 & 4.6 & 5.3 \\
& (11.3) & (11.9) & (12.0) & (9.5) & (2.4) & (10.1) & (4.0) & (0.3) & (0.4) & (3.2)
\\[3pt]
(0.5,~6) & 87.5 & 81.7 & 68.8 & 83.3 & 50.0 & 86.7 & 49.7 & 4.5 & 4.8 & 5.1 \\
& (10.7) & (9.6) & (13.5) & (6.7) & (0) & (9.5) & (2.4) & (0.3) & (0.3) & (2.8) \\
(0.5,~8) & 81 & 81.3 & 60.5 & 83.0 & 49.0 & 86.0 & 49.7 & 4.6 & 4.9 & 5.3 \\
& (11.1) & (12.8) & (11.6) & (7.9) & (4.0) & (10.3) & (2.4) & (0.3) & (0.4) & (3.5)
\\[3pt]
(0.2,~6) & 86.2 & 81.8 & 66.8 & 86.0 & 49.3 & 89.3 & 49.3 & 4.5 & 4.9 & 4.8 \\
& (10.1) & (9.2) & (12.1) & (10.3) & (3.3) & (10.5) & (3.3) & (0.3) & (0.4) & (2.2) \\
(0.2,~8) & 79.8 & 80.2 & 63.1 & 83.3 & 49.3 & 85.0 & 49.7 & 4.6 & 5.0 & 4.2 \\
& (12.6) & (14.9) & (9.2) & (8.9) & (3.3) & (9.7) & (2.4) & (0.3) & (0.3) & (2.3) \\
\hline
\end{tabular*}
\end{table}

{Among all alternative approaches, Mendel~3 has the best performance
followed by Mendel~1 and hard-thresholding-based approaches. Both
Mendels 3~and~1 test the interactions between marginally
important SNPs and all SNPs, but Mendel~3 selects marginally important
SNPs by LASSO regressions and Mendel~1 is based on the conventional
marginal analysis. Mendels 2~and~4 cannot give statistical power
greater than 50\% since only interactions among top SNPs are
considered. In terms of hard-thresholding-based TS-SIS procedures
(``Hard-SCAD'' and ``Hard-LASSO''), their performance is less
satisfactory since too many variables retained after variable screening
lead to a lower statistical power and an inflated false positive rate.
But similar to Table~\ref{tab2}, SCAD regression tends to be associated with a
higher statistical power and a smaller false positive rate. Last, the
Screen and Clean method has a low statistical power and a large and
unstable false positive rate.}

In summary, TS-SIS guided by the RATE approach is effective and
efficient in selecting truly important genetic effects and eliminating
false positives for the following penalized regressions. In the context
of the ultrahigh-dimensional GWAS model where a huge number of
potential predictors are considered, they are recommended in the real
data analysis.

\section{Framingham data analysis}\label{sec6}
We use the newly developed framework to analyze a real GWAS data set
from the Framingham Heart Study, a cardiovascular study based in
Framingham, Massachusetts, supported by the National Heart, Lung, and
Blood Institute and Boston University [\citet{DawMeaMoo51}]. Recently,
550,000 SNPs have been genotyped for the entire Framingham cohort
[\citet{Jaq07}], {from which 977 unrelated subjects including 418
males and 559 females were randomly chosen for our data analysis,
conforming to the assumption of population-based GWAS.} For each
subject, body mass index (BMI) is measured at multiple time points
between age 29 and age 61. We take the first measurement for each
individual, although the age of receiving the first measurement varies
across individuals.

As a common practice in GWAS analysis, SNPs with rare allele frequency $<$10\% were excluded from data analysis, which leaves 349,985 SNPs
across 23 chromosomes of the whole genome. {5.16\% of the remaining
SNPs, however, contain missing genotypes for some subjects. Since we
are interested in detecting active genetic effects rather than handling
missing data in this study, for each missing genotype of each subject,
we randomly draw a genotype according to the SNP's genotypic
frequencies across all subjects whose genotypes are known.} Then, by
including gender and age as two covariates, we follow the procedure
described in previous sections to select SNPs with active main effects
and construct an epistatic network explaining the observed BMI
variations. {In the RATE assisted TS-SIS procedure, in particular, the
confidence level is the same as that in simulation studies ($\beta=
0.0001$), but $\alpha$ is set to 0.0005 in screening for main effects
and to 0.00001 in detecting interactions.}

Out of 349,985 SNPs and numerous two-way interaction terms, 23 active
main effects and 24 active epistatic interactions are detected by the
TS-SIS procedure followed by SCAD penalized regression. Then, we refit
a linear regression model with these selected SNPs and two covariates
being predictors, and obtain the estimated regression coefficient and
heritability for each selected SNP. Tables~\ref{realmain} and \ref
{realinter} tabulate the information of selected SNPs with nonzero
main and epistatic interaction effects, respectively, including
chromosomes, names, minor allele frequencies (MAF), estimated genetic
effects and heritabilities. {Specifically, heritability is the
proportion of the phenotypic variance explained by the genetic variance
of a particular effect. For an additive or dominant effect, it is
calculated as}
\[
h^2 = \frac{2 p_A p_a(\hat{a}_j+(p_A-p_a)\hat{d}_j)^2+ (2p_A p_a
\hat{d}_j)^2}{\operatorname{var}(y)},
\]
{where $p_A$ is the allele frequency for $A$ and $p_a$ is the allele
frequency for $a$. For the epistatic interactions, the heritability
calculation under our general genetic model is more involved. Suppose
SNP $j$ has alleles $A$ and $a$, and SNP $j'$ has alleles $B$ and $b$.
Then for genotypes $\mathit{AABB}$, $\mathit{AABb}$, $\mathit{AAbb}$, $\mathit{AaBB}$, $\mathit{AaBb}$, $\mathit{Aabb}$,
$\mathit{aaBB}$, $\mathit{aaBb}$ and $\mathit{aabb}$, the vector of genotype frequencies is}
\begin{eqnarray*}
\bolds{\omega} &=& \bigl(p_A^2 p_B^2,
2 p_A^2 p_B p_b,
p_A^2 p_b^2, 2p_A
p_a p_B^2, 4p_A
p_a p_B p_b,
\\
&&\hspace*{75pt}2p_A
p_a p_b^2, p_a^2
p_B^2, 2 p_a^2 p_B
p_b, p_a^2 p_b^2
\bigr)^T,
\end{eqnarray*}
{and the associated genetic values are }
\begin{eqnarray*}
{\mathbf g} &=&
\bigl(\hat{a}_j+\hat{a}_{j'}+
\widehat{\mathcal{I}}_{j j'}^{aa}, \hat {d}_j+
\hat{a}_{j'}+\widehat{\mathcal{I}}_{j j'}^{da}, -\hat
{a}_j+\hat{a}_{j'}-\widehat{\mathcal{I}}_{j j'}^{aa}
\\
&&\hspace*{5pt} \hat{a}_j+\hat{d}_{j'}+\widehat{
\mathcal{I}}_{j j'}^{ad}, \hat {d}_j+
\hat{d}_{j'}+\widehat{\mathcal{I}}_{j j'}^{dd}, -\hat
{a}_j+\hat{d}_{j'}-\widehat{\mathcal{I}}_{j j'}^{ad}
\\
&&\hspace*{5pt} \hat{a}_j+\hat{a}_{j'}-\widehat{
\mathcal{I}}_{j j'}^{aa}, \hat {d}_j-
\hat{a}_{j'}-\widehat{\mathcal{I}}_{j j'}^{da}, -\hat
{a}_j-\hat{a}_{j'}+\widehat{\mathcal{I}}_{j j'}^{aa}
\bigr)^T.
\end{eqnarray*}
{Therefore, the genetic variance is $\bolds{\omega}^T {\mathbf g}^2
- (\bolds{\omega}^T {\mathbf g})^2 $, and the epistatic variance is
this genetic variance minus the genetic variances of two main effects.
Finally, the associated epistatic heritability is the epistatic
variance divided by the phenotypic variance. If dominant effects are
not modeled, this formula gives exactly the same result as the one
proposed in \citet{WuZha09}, where two SNPs' additive effects and
their additive${}\times{}$additive interaction are considered.}

%
\begin{table}
\tabcolsep=0pt
\caption{Information of SNPs with active main effects in the Framingham Heart Study}\label{realmain}\label{tab6}
\begin{tabular*}{\tablewidth}{@{\extracolsep{\fill}}@{}lccd{2.2}cd{2.2}ccd{2.2}c@{}}
\hline
\multicolumn{5}{@{}c}{\textbf{Additive effects}} &\multicolumn{5}{c@{}}{\textbf{Dominant effects}}\\[-6pt]
\multicolumn{5}{@{}c}{\hrulefill} &\multicolumn{5}{c@{}}{\hrulefill}
\\
& & & & \multicolumn{1}{c}{\textbf{Heritability}} & & & & & \multicolumn{1}{c@{}}{\textbf{Heritability}}
\\
\textbf{Chr.} & \multicolumn{1}{c}{\textbf{Name}} & \multicolumn{1}{c}{\textbf{MAF}} & \multicolumn{1}{c}{\textbf{Effect}}
& \multicolumn{1}{c}{\textbf{(\%)}} & \multicolumn{1}{c}{\textbf{Chr.}}&\multicolumn{1}{c}{\textbf{Name}}
&\multicolumn{1}{c}{\textbf{MAF}}&\multicolumn{1}{c}{\textbf{Effect}} & \multicolumn{1}{c@{}}{\textbf{(\%)}}
\\
\hline
\phantom{0}1 & ss66041272 & 0.49 & -0.82 & 1.92 & 3 & ss66173500 & 0.29 & -0.26 & 0.35 \\
\phantom{0}1 & ss66276746 & 0.13 & -0.42 & 0.23 & 3 & ss66142093 & 0.30 & -0.63 & 2.06 \\
\phantom{0}4 & ss66346559 & 0.28 & -0.51 & 0.60 & 4 & ss66354801 & 0.27 & 0.30 & 0.45 \\
\phantom{0}4 & ss66159949 & 0.29 & 0.12 & 0.03 & 6 & ss66166806 & 0.34 & -0.45 & 1.09 \\
\phantom{0}5 & ss66316662 & 0.38 & 0.37 & 0.37 & 6 & ss66299053 & 0.34 & 0.41 & 0.91 \\
\phantom{0}5 & ss66118377 & 0.50 & 0.06 & 0.01 & 6 & ss66090554 & 0.27 & 0.24 & 0.29 \\
\phantom{0}7 & ss66083530 & 0.19 & -0.47 & 0.39 & 7 & ss66083530 & 0.35 & -0.28 & 0.43 \\
\phantom{0}8 & ss66177628 & 0.23 & -0.01 & 0.00 & 7 & ss66249128 & 0.33 & -0.64 & 2.19 \\
\phantom{0}9 & ss66095597 & 0.28 & -0.60 & 0.83 & 7 & ss66314446 & 0.21 & 0.62 & 1.70 \\
12 & ss66086159 & 0.36 & -0.45 & 0.53 & 8 & ss66381612 & 0.27 & -0.32 & 0.51 \\
21 & ss66511535 & 0.16 & -0.44 & 0.30 & 11 & ss66369823 & 0.25 & 0.21 & 0.21 \\
&&& & &  13 & ss66487154 & 0.30 & 0.38 & 0.75 \\
\hline
\end{tabular*}
\end{table}

%
\begin{table}
\tabcolsep=0pt
\caption{Information of SNPs with significant interactions in the Framingham Heart Study}\label{realinter}\label{tab7}
\begin{tabular*}{\tablewidth}{@{\extracolsep{\fill}}@{}lcd{1.2}d{2.0}ccd{2.2}c@{}}
\hline
\multicolumn{3}{@{}c}{\textbf{Root 1}} & \multicolumn{3}{c}{\textbf{Root 2}}\\[-6pt]
\multicolumn{3}{@{}c}{\hrulefill} & \multicolumn{3}{c}{\hrulefill}\\
\textbf{Chr.} & \textbf{Name} & \multicolumn{1}{c}{\textbf{MAF}}
& \multicolumn{1}{c}{\textbf{Chr.}}&\multicolumn{1}{c}{\textbf{Name}}&\multicolumn{1}{c}{\textbf{MAF}} &
\multicolumn{1}{c}{\textbf{Effect}} & \multirow{2}{45pt}{\\[-34pt] \centering{\textbf{Heritability (\%)}}}\\
\hline
\multicolumn{8}{@{}c@{}}{\textit{Additive${}\times{}$additive interactions}}\\
\phantom{0}1 & ss66041272 & 0.49 & 6 & ss66061582 & 0.21 & -0.95 & 2.08 \\
\phantom{0}9 & ss66095597 & 0.28 & 4 & ss66151090 & 0.11 & -0.80 & 0.89
\\[3pt]
\multicolumn{8}{@{}c@{}}{\textit{Additive${}\times{}$dominant interactions}}\\
\phantom{0}1 & ss66137441 & 0.49 & 17 & ss66248774 & 0.49 & 1.06 & 1.70 \\
\phantom{0}3 & ss66081331 & 0.28 & 3 & ss66142093 & 0.35 & 0.56 & 0.30 \\
\phantom{0}3 & ss66375852 & 0.38 & 23 & ss66107600 & 0.33 & -1.13 & 0.83 \\
\phantom{0}4 & ss66159949 & 0.29 & 11 & ss66132273 & 0.38 & 0.74 & 0.71 \\
\phantom{0}8 & ss66177628 & 0.23 & 13 & ss66487154 & 0.34 & -1.06 & 1.32
\\[3pt]
\multicolumn{8}{@{}c@{}}{\textit{Dominant${}\times{}$additive interactions}}\\
\phantom{0}3 & ss66142093 & 0.3 & 2 & ss66430035 & 0.25 & -1.00 & 0.97 \\
\phantom{0}3 & ss66142093 & 0.3 & 3 & ss66081331 & 0.28 & -0.74 & 0.61 \\
\phantom{0}3 & ss66142093 & 0.3 & 3 & ss66483001 & 0.30 & -0.33 & 0.14 \\
\phantom{0}4 & ss66354801 & 0.27 & 7 & ss66416257 & 0.21 & -0.72 & 0.53 \\
\phantom{0}6 & ss66316737 & 0.29 & 21 & ss66113670 & 0.10 & 1.42 & 2.75 \\
\phantom{0}7 & ss66468842 & 0.33 & 7 & ss66083530 & 0.19 & -0.14 & 0.03 \\
\phantom{0}7 & ss66249128 & 0.33 & 8 & ss66047672 & 0.23 & -0.88 & 0.79 \\
15 & ss66058021 & 0.38 & 1 & ss66325411 & 0.17 & 1.01 & 0.90
\\[3pt]
\multicolumn{8}{@{}c@{}}{\textit{Dominant${}\times{}$dominant interactions}}\\
\phantom{0}3 & ss66142093 & 0.3 & 4 & ss66444506 & 0.26 & 1.09 & 0.89 \\
\phantom{0}3 & ss66142093 & 0.3 & 8 & ss66468875 & 0.39 & 0.97 & 0.71 \\
\phantom{0}3 & ss66142093 & 0.3 & 11 & ss66152909 & 0.35 & 1.08 & 0.78 \\
\phantom{0}7 & ss66468842 & 0.33 & 11 & ss66318229 & 0.29 & 1.02 & 0.68 \\
\phantom{0}7 & ss66249128 & 0.33 & 12 & ss66451087 & 0.14 & 2.25 & 2.29 \\
\phantom{0}7 & ss66249128 & 0.33 & 12 & ss66109005 & 0.16 & -1.08 & 0.61 \\
11 & ss66369823 & 0.25 & 10 & ss66482189 & 0.42 & 1.23 & 1.74 \\
11 & ss66369823 & 0.25 & 3 & ss66142093 & 0.35 & -0.40 & 0.15 \\
18 & ss66306728 & 0.3 & 16 & ss66394113 & 0.13 & 1.19 & 0.55 \\
\hline
\end{tabular*}
\end{table}

{Generally speaking, main genetic effects contribute to 16.1\% of the
phenotypic BMI variation, among which 5.2\% is due to the additive
genetic effects and 10.9\% is due to the dominant genetic effects.
Epistasis, on the other hand, explains 23.0\% of the phenotypic
variation. It is worth noting that a few SNPs and interactions
demonstrate stronger genetic effects than others. In other words,
although the expression of the BMI trait is determined by many SNPs,
there exist some SNPs that may be more influential. For example, out of
the 23 SNPs exhibiting significant additive or dominant effects, five
have heritabilities greater than 1\%. This number increases to six for
epistatic interactions.}


%
\begin{figure}[t]

\includegraphics{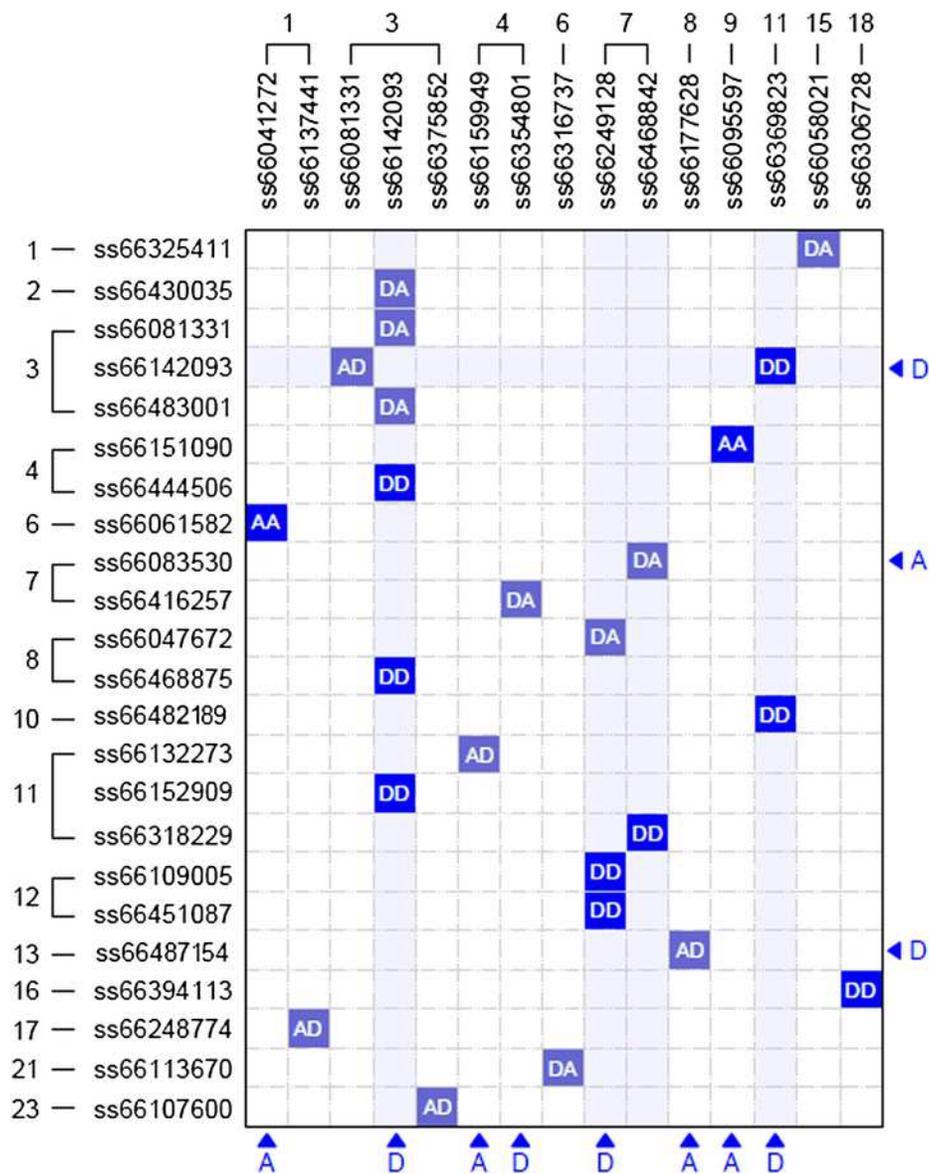}

\caption{A picture of significant SNP--SNP interactions for BMI in the
Framingham Heart Study. The numbers beside SNPs are chromosome numbers.
The SNPs that display significant additive (A) and dominant (D) effects
are indicated by arrows. The pairs of SNPs with significant additive${}\times{}$additive, additive${}\times{}$dominant, dominant${}\times{}$additive
and dominant${}\times{}$dominant interactions are indicated as
AA, AD, DA and DD, respectively.}\label{fig1}\vspace*{-7pt}
\end{figure}

To depict an overall picture of genetic control for BMI by SNP--SNP
epistasis, we draw a web of additive${}\times{}$additive, additive${}\times{}$dominant, dominant${}\times{}$additive and
dominant${}\times{}$dominant interactions in Figure~\ref{fig1} which shows the genomic distribution
of SNPs that interact with each other. From this figure, we obtain the
following interesting results: (1)~epistasis appears to be distributed
randomly throughout the genome, although a few SNPs, such as ss66142093
on chromosome 3 and ss66249128 and ss66468842 on chromosome 7 tend to
interact with many other SNPs. (2) Active epistasis may not be due to
interactions between two SNPs, both of which display active marginal
effects. Of the 24 selected pairs, there are two cases in which both
SNPs have active marginal effects and there are 14 cases in which only
one SNP has an active marginal effect, whereas the counterpart has
none. There are as many as 8 pairs in which no SNP is active for its
marginal effect. Notably, the dominant${}\times{}$dominant interaction
between SNP ss66249128 on chromosome 7 and SNP ss66451087 on chromosome
12 can explain 2.29\% of the BMI variation, although the latter is
marginally uncorrelated with BMI. In the presence of SNP ss66451087,
the dominant genetic effect of SNP ss66249128 is dramatically impacted
(Table~\ref{tab7}).

{Since our model allows a large number of SNPs to be analyzed
simultaneously, the resulting discoveries should be more biologically
relevant and statistically robust than those from traditional single-SNP approaches. For example, SNP ss66142093 on chromosome 3 was
detected to explain 2.97\% heritability. This SNP is near a candidate
gene ANAPC13 involved in pathways for bone and cartilage development
that affects human height and stature through cell cycle regulation and
mitosis [\citet{WeeFra08}].}

{In other GWAS for BMI [\citet{Fraetal07}; \citet{Scuetal07};
\citet{Speetal10}], significant SNPs were repeatedly detected on
chromosomes 1, 3, 4, 6, 7 and 11 in NEGR1, ETV5, GNPDA2. BDNF and MTCH2
loci. Our results of main genetic effects are in agreement with
previous reports about the presence of common variants near these loci
associated with biochemical pathways toward obesity. The result on
epistatic interactions suggests large epistatic effects among
chromosomes 3, 7 and 11 which have not been reported in previous
studies, possibly showing the unique power of this new approach. A
recent review on identifying genes responsible for type 2 diabetes
confirms genomic regions harboring disease susceptibility loci
[\citet{Fraetal07}]. These regions include two on chromosome 3 and
one on each of chromosomes 4, 6, 9 and 12. We have noticed many SNPs
identified in this study overlap with those detected by previous
studies targeting type 2 diabetes, suggesting the underlying
correlations between BMI and type 2 diabetes.} Additionally, our
analysis shows that the regression coefficients for gender and age are
$-$0.12 and 0.01, respectively. That is, after adjusting for these
genetic factors, the risk of obesity is higher for females, and the
risk increases with age.

{To further evaluate the significance and predictability of the
proposed method, we randomly partition the original real data set into
two parts: the training data set with 900 subjects and the validation
data set with the remaining 77 individuals. We apply the proposed RATE
assisted TS-SIS followed by the SCAD penalized regression to the
training data set, and then use the validation data set to evaluate the
estimated model. Denote by $Y_i^*$ the response BMI value of the $i$th subject in the validation data set,
and $\widehat Y_i^*$ the predicted BMI value by the estimated model
using the training data set, where $i=1,2,\ldots,77$.
We compute the following two criteria to evaluate the prediction
performance. First, we calculate the relative mean absolute prediction
error (RMAPE), which is the difference between $Y_i^*$ and $\widehat
Y_i^*$ divided by the true value of $Y_i^*$:}
\[
\mbox{RMAPE}=\frac{1}{77}\sum^{77}_{i=1}
\frac{\llvert
Y_i^*-\widehat Y_i^*\rrvert }{Y_i^*}. %
\]
{Second, we note that a primary interest of predicting BMI is to
predict whether the individual is obese or not, that is, BMI$>30$. Thus,
we compute the classification accuracy (CA) of the validation data set
using the estimated model:}
\[
\mbox{CA}=1-\frac{1}{77}\sum^{77}_{i=1}
\bigl\llvert I\bigl(Y_i^*>30\bigr)-I\bigl(\widehat
Y_i^*>30\bigr)\bigr\rrvert, %
\]
{where $I(\cdot)$ is an indicator function. Then, we repeat the above
validation experiment 10 times. The average RMAPE is 14.10\%, and the
standard deviation of RMAPE is 0.85\%. The average CA is 82.77\%, with
a standard deviation of 3.37\%. These results suggest that our model
predicts well in the out-of-sample validation data sets.}

\section{Discussion}\label{sec7}
Identifying genetic interaction network is an important task in
genome-wide association studies, but is challenged by the sheer volume
of genetic data. In this paper we present a comprehensive GWAS model
and propose a statistical framework to identify important SNPs and
interactions which jointly explain the observed phenotypes.
Specifically, a two-stage sure independence screening procedure
(TS-SIS) is proposed to formulate a candidate pool of SNPs, including
those without weak main effects, but serving as a root in two-way
interactions. This procedure expands the literature by relaxing the
restrictive assumption that two roots in an interaction have to be
marginally correlated with the response. A~RATE approach is also
proposed to determine the number of predictors retained in each stage
of TS-SIS. This approach can also be applied to other variable
screening problems.

{\citet{WuZha09} derived an analytical approach to calculate the
power of a model selection strategy in GWAS that is similar to the
proposed TS-SIS. Their approach allows for random genotypes,
correlation among test statistics as well as a false-positive control.
It is straightforward to apply their power calculations to our
framework. } Since the TS-SIS procedure provided a relatively
low-dimensional regression model containing important SNPs with high
probability, existing penalized least squares estimations and their
empirical performances in GWAS analysis provided valuable guidance for
selecting important SNPs and constructing a gene--gene interaction network.

The new model has been used to analyze GWAS data from the Framingham
Heart Study [\citet{DawMeaMoo51}], aimed to identify genetic variants
that affect cardiovascular diseases and their related traits such as
blood pressure and BMI [\citet{Jaq07}]. To the best of our knowledge,
this is likely the first study that has detected genetic interactions
for obesity-related traits in GWAS. Since the detected SNPs displaying
important interactions may be harbored in genes of the BMI-associated
metabolic pathways [\citet{Speetal10}], plus higher
heritabilities collectively explained by them, our model should provide
a powerful and useful tool for understanding the underlying genetic
mechanisms and regulatory network of obesity. For example, dopamine,
which is a neurotransmitter, modulates motivation and rewarding
properties of eating. \citet{Wanetal01} confirmed by biomedical
experiments that brain dopamine levels are significantly lower in the
obese individuals, suggesting strong correlations between BMI and
genetic regulatory networks. The use of our model to detect
dopamine-associated SNPs in a GWAS study should help to unravel the
genetic architecture of obesity.

Our statistical procedure is capable of identifying epistatic
interactions and enables researchers to decipher a detailed picture of
the genetic architecture of human diseases or complex traits. So far,
we have concentrated on detecting interactions for a continuous trait
in GWAS. The proposed TS-SIS assisted SCAD regression can be readily
extended to case-control cohorts, family trios or survival data
analysis in genome-wide association studies. The framework can also be
applied to other statistical problems, where the accurate detection of
interactions is desired in the presence of high-dimensional data sets
or ultrahigh-dimensional data sets.

\begin{appendix}
\section*{Appendix}

\begin{pf*}{Proof of Theorem \protect\ref{teofpr}}
Let any $r\in\mathcal{N}_+$, the set of positive integers. The event
$\{\llvert \widehat{\mathcal{M}}\cap\mathcal{M}^c\rrvert
\ge r\}$ represents that at least $r$ unimportant variables rank on the
top of all auxiliary variables. Because the inactive variables $\{X_j\dvtx
j\in\mathcal{M}^c\}$
and auxiliary variables $\{Z_k\dvtx  k=1,\ldots,d\}$ are exchangeable, we
follow the idea of \citet{Zhuetal11} and have that
%
\begin{eqnarray}\label{fpr1}
\qquad P \bigl(\bigl\llvert \widehat{\mathcal{M}}\cap
\mathcal{M}^c\bigr\rrvert \ge r \bigr) &\leq& \frac{(p_n-s_n)!}{(p_n-s_n-r)!r!} \Big/
\frac{(p_n-s_n+d)!}{(p_n-s_n+d-r)!r!}\nonumber
\\
&\leq& \frac{(p_n-s_n+d-r)\times\cdots\times
(p_n-s_n+1-r)}{(p_n-s_n+d)\times\cdots\times(p_n-s_n+1)}
\\
&\leq& \biggl(1-\frac{r}{p_n+d} \biggr)^d.\nonumber
\end{eqnarray}
$\llvert \mathcal{M}^c\rrvert =p_n-s_n>p_n-n$ by the sparsity
principle. If we can assume $\llvert \mathcal{M}\rrvert <n$,
it~follows that
%
\begin{eqnarray}
\label{fpr2}
P \biggl(\frac{\llvert \widehat{\mathcal{M}}\cap\mathcal
{M}^c\rrvert }{\llvert \mathcal{M}^c\rrvert }<\alpha \biggr) &=& 1- P
\bigl(\bigl\llvert \widehat{\mathcal{M}}\cap\mathcal {M}^c\bigr\rrvert
\ge\alpha\bigl\llvert \mathcal{M}^c\bigr\rrvert \bigr)\nonumber
\\
&\ge& 1- P \bigl(\bigl\llvert \widehat{\mathcal{M}}\cap\mathcal {M}^c
\bigr\rrvert \ge\alpha(p_n-n) \bigr)
\\
&\ge& 1- \biggl\{1-
\frac{\alpha(p_n-n)}{p_n+d} \biggr\}^d,\nonumber
\end{eqnarray}
where the second inequality follows by (\ref{fpr1}). 
\end{pf*}
\end{appendix}

\section*{Acknowledgments}
The Framingham Heart Study project is
conducted and supported by the National Heart, Lung, and Blood
Institute (NHLBI) in collaboration with Boston University (N01 HC25195).

The authors acknowledge the investigators that contributed the
phenotype, genotype and simulated data for this study. The manuscript
was not prepared in collaboration with investigators of the Framingham
Heart Study and does not necessarily reflect the opinions or views of
the Framingham Heart Study, Boston University or the NHLBI.
The authors are grateful to Dr. Zhong Wang for sharing the idea to
create Figure~\ref{fig1} in this paper.
The authors thank the Editor, the Associate\vadjust{\goodbreak} Editor and three anonymous referees for
their constructive comments,
which have led to a significant improvement of the earlier version of
this paper.
The content is solely the responsibility of the
authors and does not necessarily represent the official views of the
NIDA or NNSFC.


%

\printaddresses
\end{document}